\pdfoutput=1
\documentclass[11pt]{iopart}

\usepackage{amssymb}
\usepackage{graphics}
\usepackage{graphicx}
\usepackage[usenames,dvipsnames]{color}

\begin{document}
\title[Efficient Guiding of Cold Atoms though a Photonic Band Gab Fiber]{Efficient Guiding of Cold Atoms though a Photonic Band Gab Fiber}%
\author{S.~Vorrath, S.~A.~M\"oller, P. Windpassinger, K.~Bongs\footnote{current address: MURAC, School of Physics and Astronomy, University of Birmingham, Edgbaston, Birmingham B15 2 TT, UK}, and K.~Sengstock}
\address{Institute for Laser Physics, University of Hamburg, Luruper Chaussee 149, 22761 Hamburg, Germany}%
\ead{sengstock@physik.uni-hamburg.de}
\begin{abstract}We demonstrate the first guiding of cold atoms through a 88\,mm long piece of photonic band gap fiber. The guiding potential is created by a far-off resonance dipole trap propagating inside the fiber with a hollow core of 12\,$\mu m$. We load the fiber from a dark spot $^{85}$Rb magneto optical trap and observe a peak flux of more than $10^5$\,atoms/s at a velocity of 1.5\,m/s. With an additional reservoir optical dipole trap, a constant atomic flux of $1.5\times10^4$\,atoms/s is sustained for more than 150\,ms. These results open up interesting possibilities to study nonlinear light-matter interaction in a nearly one-dimensional geometry and pave the way for guided matter wave interferometry.
\end{abstract}
%

\maketitle

\section{Introduction}\label{intro}

The generation of Bose-Einstein condensates \cite{Anderson_Science_1995, Davis_PRL_1995}  has equipped modern atomic physics with an extremely versatile tool to study coherent atomic matter waves. Guides for matter waves have been proposed and realized using both magnetic \cite{Wang_PRL_2005, Jones_PRL_2003, Muller_PRL_1999} and optical \cite{Ito_PRL_2006, Renn_PRA_1997, Muller_PRA_2000} potentials. Similarly to an optical fiber, an ideal matter waveguide can be used to transport a single spatial mode of a coherent matter wave source, thus enabling matter wave interferometry \cite{Badurek_Book_1988, Berman_Book_1997, Andrews_Science_1997, Schumm_NatPhys_2005} over large distances. To this end, cold atoms need to be transported in a lossless, tightly confining potential in order to preserve the coherence properties of the matter wave. Extensive studies have been performed both theoretically \cite{Ol'Shanii_OpticsComm_1993, Marksteiner_PRA_1994, Ito_OpticsComm_1995, Harris_PRA_1995} and experimentally \cite{Renn_PRL_1995, Renn_PRA_1996, Dall_JOptB_1999, Muller_PRA_2000, Tarasishi_QuantumElec_2000}  using different geometries of hollow glass capillaries and exploiting the confining potentials created by the ac-Stark shift of atoms in a detuned laser field. However, severe losses of the confining light, speckle patterns and multimode performance limited the useful guiding length. These shortcomings can be overcome by exploiting the specific features of the recently developed hollow-core photonic band gap (HCPBG) fibers which do not suffer from bending losses and speckle \cite{Russell_Science_2003}. These fibers can be produced to support single-mode potentials at different wavelengths. On the one hand, this allows for matter wave guiding in high power, large red detuned potentials with low spontaneous scattering. On the other hand, at detunings closer to the atomic resonances, low-light-level strong non-linear interactions can be studied \cite{Chang_NatPhys_2008,Kiffner_PRA_2010}. Recently, thermal atoms have been guided through  \cite{Takekoshi_PRL_2007} and cold atoms have been moved into \cite{Christensen_PRA_2008} a HCPBG fiber. Interactions of thermal \cite{Benabid_Science_2002, Konorov_OptLett_2003, Gosh_PRL_2006} and cold atoms \cite{Bajcsy_PRL_2009} have been observed in HCPGB fibers,  but so far, guiding of \textit{cold, slow atoms through}
such a fiber has remained an elusive goal. In this manuscript, we report on the first observation of an efficient matter waveguide for cold, slow atoms.
\begin{figure}[h!]
\includegraphics[width=\columnwidth]{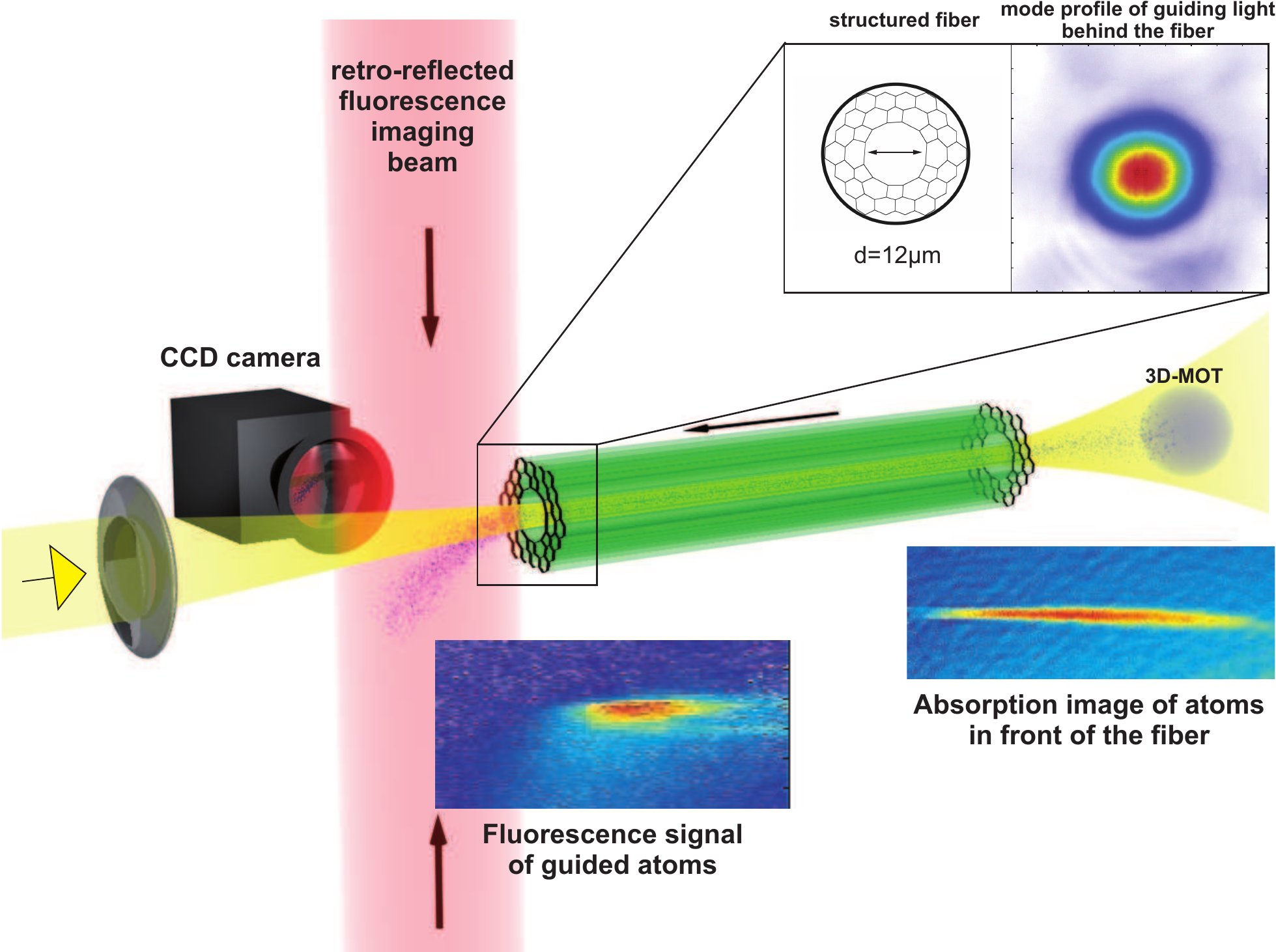}
\caption{Schematic drawing of the experimental setup. A cloud of laser-cooled atoms is prepared in front of the photonic band gap fiber and moved into the trapping region of the fiber trap by changing the position of the magneto-optical trap with an additional  magnetic field. After transport through the fiber, the atoms are detected by fluorescence imaging with a retro-reflected laser beam and collecting the photons with an intensified CCD camera. For optimization purposes, absorption imaging can be performed of the atoms trapped by the guiding light exiting at the atom-input side of the fiber.
}\label{setup}
\end{figure}

\section{Experimental setup}
The basic principle of the experimental setup is depicted in Fig. \ref{setup}; the details of the surrounding setup have been omitted.
To realize a guiding potential inside a hollow-core photonic band-gap fiber\footnote{Crystal Fibre, Air-12-1060, core diameter $12\,\mu$m, 88\,mm long} for cold atomic rubidium, a far red detuned optical guiding potential is applied. In a far red detuned optical light field, atoms are attracted towards the regions of highest electric field which in our case is the maximum of the almost Gaussian mode inside the hollow core of the fiber. To reduce loss of atoms due to spontaneous photon scattering in the guiding field, we choose a wavelength of $\lambda=1067$\,nm. We couple two linear, orthogonally polarized modes of up to 2.3\,W power each and take special care that only the fundamental mode is addressed. While we observe that only one of the two polarization modes contributes to the guiding process itself, the second polarization mode still creates an optical potential at the cold atom input side of the fiber.
This leads to an increase of cold atoms at the fiber tip but not to an increase of the atomic flux or the transported atom number.
Furthermore we observe that the not-guiding polarization mode has no negative impact to the guiding process itself. This behavior could not be explained by now.
\begin{figure*}[t!]
\includegraphics[width=\textwidth]{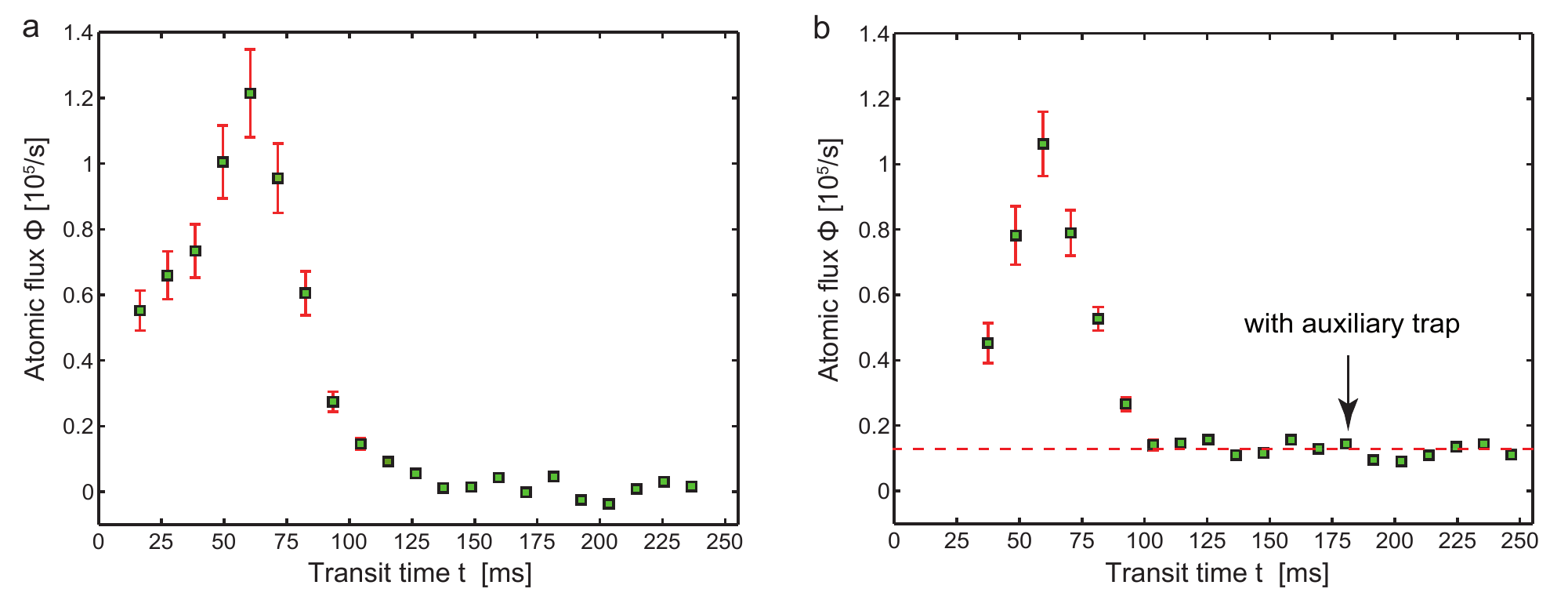}
\caption{Observed atomic flux through the hollow-core photonic band gap fiber as function of time after loading atoms into the guiding potential. \textbf{a} When atoms are loaded directly from the optical molasses, we obtain a peak flux of $(1.2\pm 0.1)\times10^5$\,atoms/s  extended over 50\,ms. \textbf{b} With an auxiliary reservoir dipole trap at the fiber input, the peak flux remains unaltered but a constant flux of $1.5\times10^4$ atoms is maintained for $\gtrsim 150$\,ms.
}\label{guiding_1}
\end{figure*}
With these parameters, a potential depth of up to 8.2\,mK can be obtained. An input guiding light coupling efficiency of over 80\% is achieved by placing the final focussing lens inside the vacuum chamber\footnote{asphere with focal length  11\,mm, numerical aperture 0.25}. As source for the cold atomic sample we use a three-dimensional magneto-optical trap (MOT)  at one end of the  HCPBG fiber, which is efficiently loaded with a cold beam from a two-dimensional MOT in a separate vacuum chamber. By adiabatically ramping up a homogeneous magnetic field during the MOT phase, the cloud is moved close to the fiber tip and during a subsequent dark spot MOT and dark spot molasses phase, a density of $1.5\times 10^{11}$\,atoms/cm$^3$ and a temperature of $10\,\mu$K are obtained. These atoms can be efficiently transferred into the guiding potential. After transport through the fiber, the guided atoms are detected on the opposite side by fluorescence imaging. The necessary sensitivity is obtained by using an intensified CCD camera which integrates the signal for 11\,ms. To balance the radiation pressure during resonant illumination we choose a retro-reflected beam arrangement. In addition, this leads to a significant increase of the number of scattered photons due to the quasi frozen dynamics in one dimension. From the efficiency of the detection setup, we estimate to detect $41\pm5$ photons per atom in the detection volume.
After loading  the cold atoms into the guiding potential of the fiber by switching off the optical molasses beams, we anticipate to observe a sharply peaked flux due to the transient nature of the loading process. This is confirmed by the experimental data shown in Fig.~\ref{guiding_1}a. The maximum of the flux is observed 60\,ms after switchoff of the molasses phase, translating into a longitudinal guiding velocity of 1.5\,m/s which corresponds very well to the guiding potential depth of 8.2\,mK. We measure a peak flux of $(1.2\pm 0.2)\times10^5$\,atoms/s and an integrated number of guided atoms of  $(7.4\pm0.3)\times 10^3$.

A quasi continuous matter wave guide can be realized by adding a long-lived reservoir trap at the cold atom input side of the fiber. To this end, we implement a second dipole trap with a beam waist of $27\,\mu$m and a trap depth of 2.2\,mK  ($\lambda=782\,$nm) which has its maximum potential depth close to the fiber tip.  Without the fiber trap engaged, we observe a maximum number of $1.8\times10^6$ atoms in the auxiliary trap at temperatures of $38\,\mu$K. Fig.~\ref{guiding_1}b shows the observed atomic flux when both traps are operated simultaneously. Again, we observe a sharply peaked temporal spectrum at the output of the fiber, resulting from guiding of atoms produced in the optical molasses. However, a nearly constant flux of $1.5\times10^4$\,atoms/s is sustained for more than 150\,ms after the maximum. To achieve this quasi continuous operation, the position of the reservoir trap's waist has to be carefully adjusted in order not to create potential barriers at the fiber input due to reflections from the fiber tip.
\begin{figure*}[ht!]
\includegraphics[width=\textwidth]{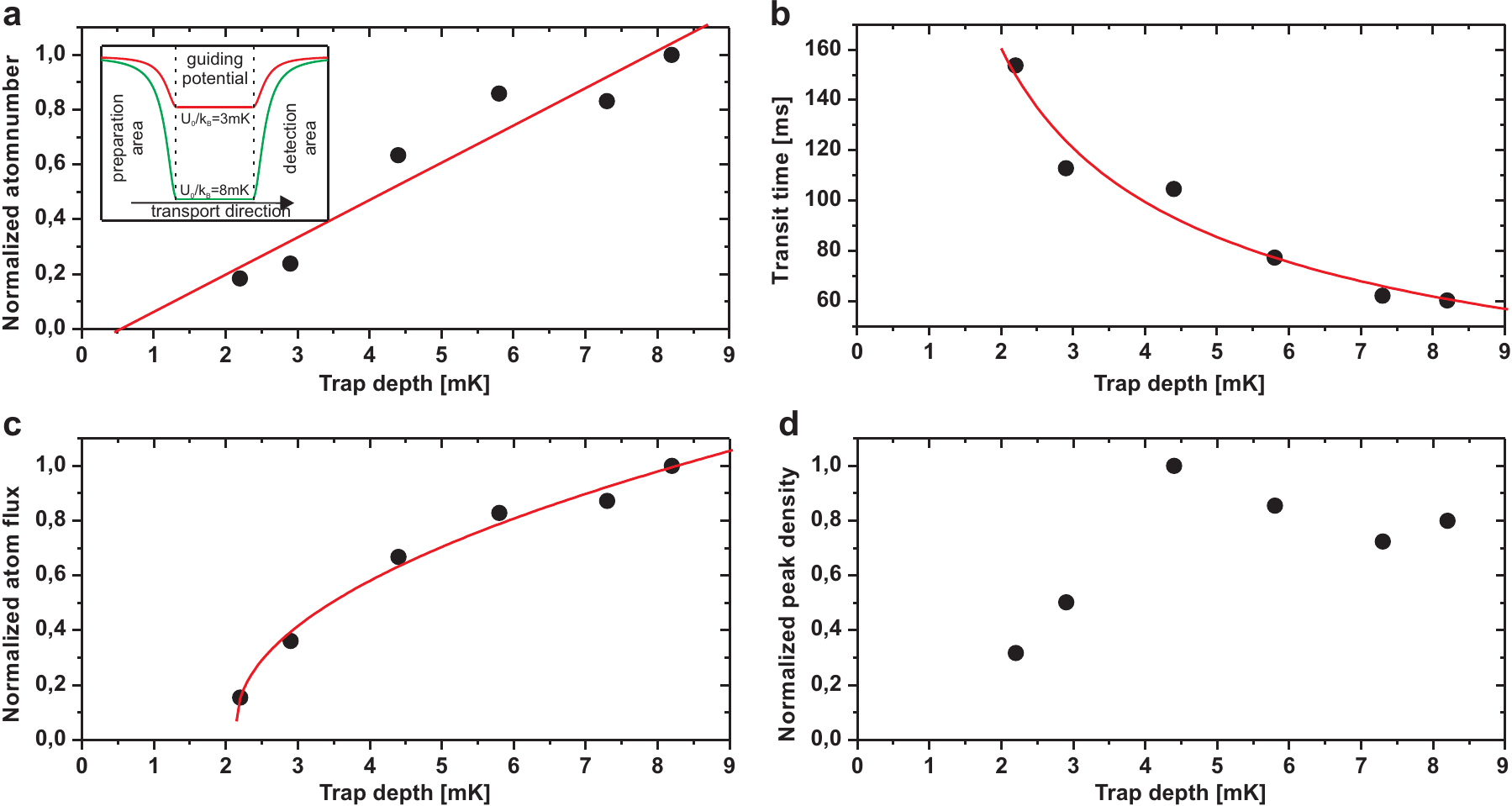}
\caption{Properties of the atomic waveguide. \textbf{a} Number of atoms guided through the fiber as a function of the guiding potential depth. The solid line represents a linear fit to the data. \textbf{Inset} Schematic drawing of the potential along the fiber axis for two different trapping depths.   \textbf{b} Transit time of the atoms through the fiber. The theoretically expected $t_\mathrm{Fly} \propto 1/\sqrt{U_0}$ fit is shown as a solid curve. \textbf{c} Atomic flux. For conservative guiding potentials, the flux is expected to scale as  $\Phi \propto \sqrt{U_0}$ as confirmed by the fit to the data (solid line). \textbf{d} Normalized peak atomic density inside the fiber.
}\label{guiding_2}
\end{figure*}

To gain deeper insight into the guiding process, we discuss the properties of the guiding potential.  In radial direction it has a nearly Gaussian shape (compare Fig.~\ref{setup}) and it can be assumed to be constant in axial direction inside the fiber core. At both ends of the fiber, the light field propagates according to classical optics and has a funnel-like shape. The inset of figure~\ref{guiding_2}a illustrates the situation exemplarily for two different guiding beam powers.

As the spontaneous scattering rate of atoms inside the fiber ($\Gamma_\mathrm{sc, max}\backsimeq 120$\,Hz) plays a minor role, atoms are expected to slide into the guiding potential, travel  through  the fiber and escape on the other side. The trapping volume for atoms at the input side of the fiber increases with the power of the trapping beam and we can expect that the number of guided atoms rises with the maximal potential depth. This  is confirmed by the experimental data as shown in Fig. \ref{guiding_2}a.  \\

From the simple picture of the guiding process (inset of Fig. \ref{guiding_2}a), we expect that the atoms gain a kinetic energy corresponding to the potential depth $U_0$ as they enter the fiber. Since their initial kinetic energy can be neglected, we thus expect the transit time through the fiber to scale as $t_\mathrm{Fly}\propto 1/\sqrt{U_0}$. The experimental data, Fig. \ref{guiding_2}b clearly exhibits this structure. As a consequence, we can assume that the potential inside the fiber is smooth and flat, which is an important prerequisite for matter wave interferometry and further applications.
\\
Additional information on the guiding characteristics can be obtained by extracting the atomic flux through the fiber. It has been shown that it is expected to scale as $\Phi \propto \sqrt{U_0}$ for a conservative potential inside a straight fiber \cite{Renn_PRA_1997}. The experimental data depicted in Fig. \ref{guiding_2}c clearly follows this functional form. However, a non-trivial threshold of $U_{0,\mathrm{min}}\approx 2.1$\,mK is observed, which is attributed to irregularities in the potential at the tip of the fiber due to reflections of the guiding light at the front surface. As a consequence, only samples with a minimal longitudinal velocity of $\langle v_\mathrm{long.}\rangle_{min}=0.45$\,m/s can be produced. This might be improved by optimizing the two facets of the fiber piece.\\
An important factor in the characterization of the atomic sample inside the fiber, especially when considering its application for non-linear optics, is the atomic density and resulting longitudinal optical depth.
An estimate can be obtained by considering the radial distribution of atoms inside the fiber. With initial temperatures of the order of $10\,\mu$K of the atoms inside a Gaussian trapping potential of $\approx 8$\,mK total depth, the atoms only occupy the innermost part of the trap. With a maximum number of $7\times10^3$ atoms being simultaneously inside the fiber, we calculate a density of $\sim 5\times 10^{11}$\,atoms\,cm$^{-3}$.
This  peak density, Fig. \ref{guiding_2}d,  shows a rise followed by a slow decay with increasing guiding potential. This behavior is explained by the two counteracting trends: As discussed, the total guided atom number increases with the potential depth.  On the other hand, the atomic ensemble is spread over a larger longitudinal region during acceleration into the guide. At low potentials the increase in atom number dominates, while at higher potentials the spreading tends to overcompensate this gain. As a consequence, experiments aiming at high in-guide density should be carefully tuned to the resulting maximum. For this maximum, we calculate an on resonance optical depth of $\sim 10^4$. In the current setup this optical depth cannot be observed directly as the HCPBG fiber does not support modes close to the D-line doublet at 780\,nm and 795\,nm, respectively, due to its band gap.

\section{Summary}
In conclusion, we have demonstrated an efficient optical guide for cold atoms. Both the transient loading regime and a quasi continuous operation of the matter wave guide have been studied. The properties of the guided atoms can be well explained by the nature of the guiding potential. The findings clearly demonstrate the feasibility of long distance transport of cold atoms and open up a whole new variety of cold atom experiments. Combined with new cooling techniques, singlemode operation of the waveguide should be achievable, which would pave the way towards guided matter wave interferometry or continuous atom laser systems \cite{Hagley_Science_1999}. As the band gap of the HCPBG fiber can be adapted to support light fields at other wavelengths than the one used here \cite{Light2009}. Thus the combination of a strong guiding potential with low spontaneous photon scattering and close-to-resonance optical driving of the atomic levels becomes feasible. Additionally, the extreme radial-to-axial size ratio of the guiding potential allows for the generation of quasi one dimensional quantum gases. In combination with a one-dimensional optical lattice inside the fiber, even strongly correlated, one dimensional systems could be studied.

\section{Acknowledgements}
This work was supported by DFG grant SE/05-247 and the DFG graduate school GrK1355.\\ \ \\

\bibliographystyle{iopart-num}

\providecommand{\newblock}{}

\end{document}